\documentclass[runningheads]{svmult}

\usepackage{makeidx}   
\usepackage{graphicx}  
\usepackage{subeqnar}  
\usepackage{multicol}  
\usepackage{physprbb}  
\makeindex             

\def\simgt{\lower 2pt \hbox{$\, \buildrel {\scriptstyle >}\over {\scriptstyle \sim}\,$}}
\def\simlt{\lower 2pt \hbox{$\, \buildrel {\scriptstyle <}\over {\scriptstyle \sim}\,$}}
\def\chandra{{\it Chandra\/}}
\def\conx{{\it Constellation-X\/}}
\def\genx{{\it Generation-X\/}}
\def\rosat{{\it ROSAT\/}}
\def\xeus{{\it XEUS\/}}
\def\xmm{{\it XMM-Newton\/}}
\def\aox{{$\alpha_{\rm ox}$}}

\begin{document}
\title*{X-rays from the First Massive Black Holes}
\toctitle{X-rays from the First Massive Black Holes}
%
%
\titlerunning{X-rays from the First Massive Black Holes}
%
\author{W.N. Brandt\inst{1}
\and C. Vignali\inst{2} 
\and B.D. Lehmer\inst{1}
\and L.A. Lopez\inst{1}
\and D.P. Schneider\inst{1}
\and I.V. Strateva\inst{1}}
\authorrunning{W.N. Brandt et al.}
%
%
\institute{Department of Astronomy \& Astrophysics, The Pennsylvania 
State University, 525 Davey Lab, University Park, PA 16802, USA
\and 
Dipartimento di Astronomia, Universit\`a degli Studi di Bologna, 
Via Ranzani 1, 40127 Bologna, Italy}

\maketitle              


\begin{abstract}
We briefly review some recent results from \chandra\ and \xmm\ studies 
of the highest redshift (\hbox{$z>4$}) active galactic nuclei (AGNs). 
Specific topics covered include radio-quiet quasars, radio-loud quasars, 
moderate-luminosity AGNs in \hbox{X-ray} surveys, and future 
prospects. No significant changes in AGN \hbox{X-ray} emission properties have 
yet been found at high redshift, indicating that the small-scale \hbox{X-ray} 
emission regions of AGNs are insensitive to the dramatic changes on larger 
scales that occur from \hbox{$z\approx 0$--6}. \hbox{X-ray} observations 
are also constraining the environments of high-redshift AGNs, relevant 
emission processes, and high-redshift AGN demography. 
\end{abstract}


\section{Introduction}

Understanding of the \hbox{X-ray} emission 
from the highest redshift (\hbox{$z>4$}) active
galactic nuclei (AGNs) has advanced rapidly over the past five years, due to 
the superb capabilities of \chandra\ and \xmm\ combined with plentiful 
high-redshift discoveries by wide-field optical surveys. The number of \hbox{X-ray} 
detections at \hbox{$z>4$} has increased from 6 in 2000 to $\approx 100$ today.

{\it Are the first massive black holes feeding and growing in the same way
as local ones?\/} \hbox{X-ray} observations can address this question effectively, 
as they probe the immediate vicinity of the black hole where changes in the
accretion mode should be most apparent. For example, high-redshift AGNs might
typically have different values of $\dot M/\dot M_{\rm Edd}$, given that they 
reside in young, forming galaxies that differ greatly from the majority of 
those in the local universe. Galactic black holes and local AGNs both show 
strong \hbox{X-ray} spectral changes with $\dot M/\dot M_{\rm Edd}$ that should 
be detectable at high redshift (e.g., Brandt 1999; McClintock \& Remillard 2004). 
\hbox{X-ray} observations can also constrain 
the environments of high-redshift AGNs 
(e.g., via \hbox{X-ray} absorption studies), 
relevant emission processes (e.g., in AGN jets), and 
high-redshift AGN demography. 

We have been addressing the issues above using a combination of 
snapshot \chandra\ observations, 
spectroscopic \xmm\ observations, 
\hbox{X-ray} survey data, and 
archival \hbox{X-ray} data. 
Below we will briefly review some of our results and will
discuss some future prospects. 
We adopt 
\hbox{$H_0=70$~km~s$^{-1}$~Mpc$^{-1}$}, 
\hbox{$\Omega_{\rm M}=0.3$}, and 
\hbox{$\Omega_{\Lambda}=0.7$} 
throughout. 


\begin{figure}[t!]
\begin{center}
\includegraphics[width=0.7\textwidth,angle=0]{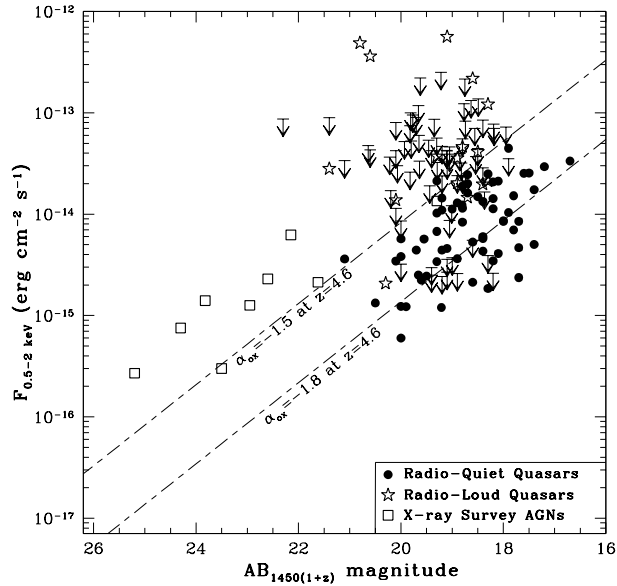}
\end{center}
\caption[]{Observed-frame, Galactic absorption-corrected \hbox{0.5--2~keV} 
flux versus AB$_{1450(1+z)}$ magnitude for $z>4$ AGNs. Object types 
are as shown in the legend; downward-pointing arrows indicate \hbox{X-ray} upper
limits. The slanted lines show \hbox{$\alpha_{\rm ox}=-1.5$} and 
\hbox{$\alpha_{\rm ox}=-1.8$} loci at a fiducial redshift of 
\hbox{$z=4.6$}. Adapted from Vignali et~al. (2005).}
\label{eps1}
\end{figure}


\section{High-Redshift Radio-Quiet Quasars}

Most of the \hbox{$z>4$} AGNs with \hbox{X-ray} detections are luminous ($M_{\rm B}<-26$) 
radio-quiet quasars (RQQs) that have been discovered in wide-field optical 
surveys and targeted with snapshot (4--12~ks) \chandra\ observations. For example, 
we have been targeting the most luminous RQQs ($M_{\rm B}\approx -27$ to $-29.5$) 
from the Palomar Digital Sky Survey (DPOSS) and the Automatic Plate Measuring 
facility survey (APM) as well as some of the highest redshift RQQs ($z>4.8$) 
from the Sloan Digital Sky Survey (SDSS). 

Figure~1 shows the \hbox{X-ray} and optical fluxes for \hbox{$z>4$} AGNs with sensitive
\hbox{X-ray} observations. As expected, the \hbox{X-ray} and optical fluxes for the RQQs 
are generally correlated, but there is significant scatter in this
relation (see Vignali et~al. 2003ac, 2005 for details). Even the
optically brightest RQQs at \hbox{$z>4$} have relatively low \hbox{0.5--2~keV} fluxes
of a few $\times 10^{-14}$~erg~cm$^{-2}$~s$^{-1}$, so \hbox{X-ray} spectroscopy
of these objects is challenging. 
Nevertheless, there has been respectable progress constraining the basic 
\hbox{X-ray} spectral properties of \hbox{$z>4$} RQQs 
using joint \hbox{X-ray} spectral fitting 
(Vignali et~al. 2003ac, 2005) and single-object spectroscopy 
(Ferrero \& Brinkmann 2003; Farrah et~al. 2004; Grupe et~al. 2004; 
Schwartz \& Virani 2004). Figure~2, for example, shows 
the results from joint \hbox{X-ray} spectral fitting 
of 48 RQQs from \hbox{$z=3.99$--6.28} (with a median redshift
of 4.43) using the Cash (1979) statistic (Vignali et~al. 2005). 
This joint-fitting approach provides a stable estimate of average 
\hbox{2--40~keV} rest-frame spectral properties. The spectrum is well 
fit by a power-law model with $\Gamma=1.93^{+0.10}_{-0.09}$ and Galactic 
absorption; any widespread intrinsic absorption by neutral material has 
\hbox{$N_{\rm H}\simlt 5\times 10^{21}$~cm$^{-2}$}. 
No iron~K line emission or Compton reflection is detected, although the
statistical constraints on such spectral components are not yet
particularly meaningful.  


\begin{figure}[t!]
\begin{center}
\includegraphics[width=0.8\textwidth,angle=0]{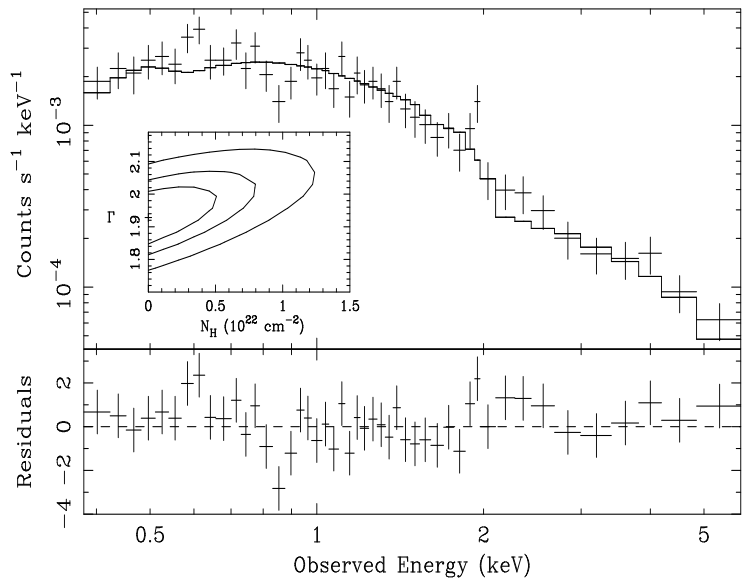}
\end{center}
\caption[]{Stacked \chandra\ spectrum constructed from 48 RQQs at 
\hbox{$z=3.99$--6.28}. The spectrum has $\approx 870$ source counts 
and an effective exposure time of 244~ks. The spectrum has been fit with a 
power-law model including Galactic absorption, and data-to-model 
residuals (in units of $\sigma$) are shown in the lower panel. The 
fit is statistically acceptable. The inset shows the 68, 90, and 99\% 
confidence regions for the power-law photon index and intrinsic column
density derived from joint spectral fitting with the Cash (1979) 
statistic. Adapted from Vignali et~al. (2005).}
\label{eps2}
\end{figure}


\begin{figure}[t!]
\begin{center}
\includegraphics[width=0.8\textwidth,angle=0]{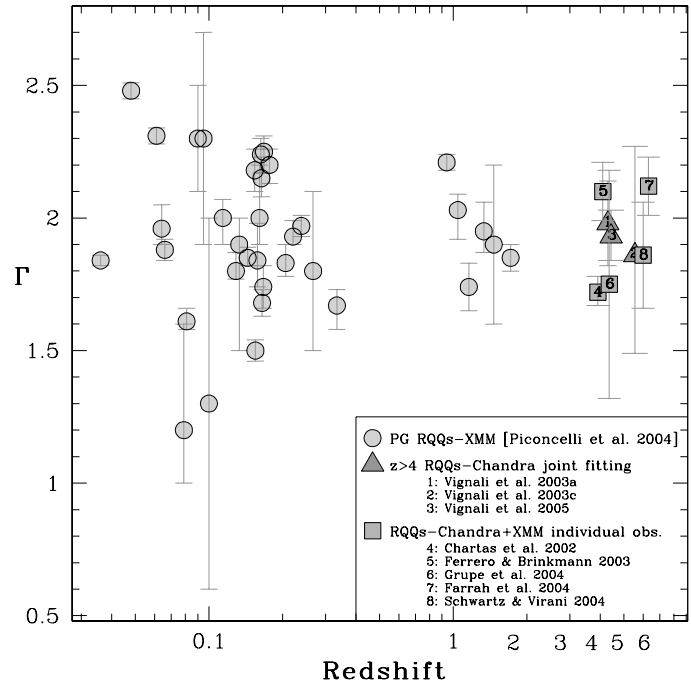}
\end{center}
\caption[]{Hard \hbox{X-ray} photon index versus redshift. The circles at 
\hbox{$z<2$} indicate RQQs from the Bright Quasar Survey (PG) analyzed by 
Piconcelli et~al. (2004). At higher redshift, the data points have been 
derived from both joint spectral fitting (triangles) and single-object 
spectroscopy (squares); the data points are numbered with corresponding 
citations given in the figure legend.}
\label{eps3}
\end{figure}


Figure~3 summarizes the current \hbox{X-ray} spectral results on high-redshift
RQQs, comparing their power-law photon indices to those of RQQs at
lower redshift. There is significant intrinsic scatter in the photon 
indices at all redshifts. This scatter is believed to be due to 
object-to-object variations in the temperature and optical depth of 
the accretion-disk corona, likely arising from variations in 
$\dot M/\dot M_{\rm Edd}$ (see Section~1). However, there is
no detectable systematic change in the photon-index distribution 
at high redshift. On average, the accretion-disk coronae of high-redshift 
and low-redshift RQQs appear to have similar properties. 


\begin{figure}[t!]
\begin{center}
\includegraphics[width=1.0\textwidth,angle=0]{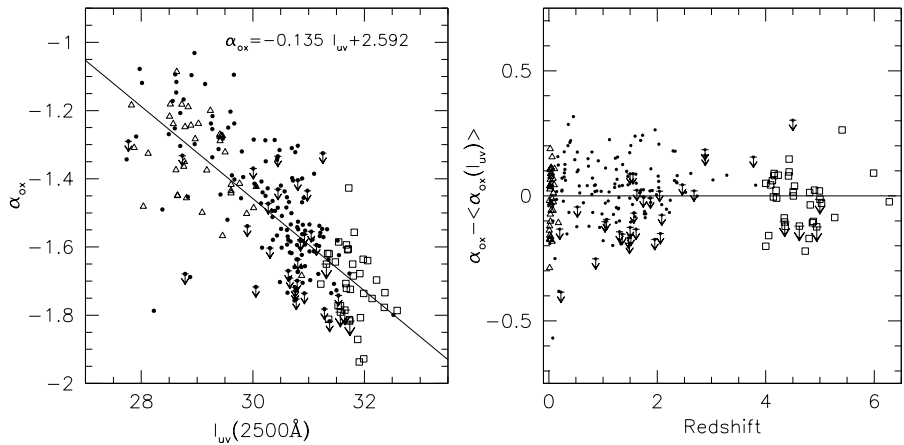}
\end{center}
\caption[]{The left panel shows \aox\ versus 2500~\AA\ luminosity 
density for optically selected, radio-quiet AGNs: 
Seyfert~1 galaxies from Walter \& Fink (1993; open triangles), 
SDSS AGNs in medium-depth \rosat\ observations (filled circles), and 
$z>4$ RQQs from Vignali et~al. (2003c; open squares). Large negative 
values of \aox\ correspond to relatively weak \hbox{X-ray} emission, 
and downward-pointing arrows indicate X-ray upper limits. 
The best fit to the data is also shown. 
The right panel shows \aox\ residuals versus redshift after removing 
the best-fit luminosity dependence of \aox. Note that no systematic
trends remain. 
Adapted from Strateva et~al. (2005).}
\label{eps4}
\end{figure}


In addition to measuring \hbox{X-ray} spectral properties, it is also informative
to investigate relations between \hbox{X-ray} and longer wavelength 
emission. Any changes in accretion mode over cosmic time might lead 
to changes in the fraction of total power emitted as X-rays. Investigations 
of this type have been performed since the 1980's
(e.g., Avni \& Tananbaum 1986), often utilizing \aox,  
which is defined as the point-to-point spectral slope between 2500~\AA\ 
and 2~keV in the rest frame. Studies of the dependence of \aox\ upon
redshift and luminosity are challenging for several reasons: 
(1) broad ranges of sample redshift and luminosity are required to 
break statistical degeneracies between these two quantities, since
they are correlated in flux-limited AGN samples, 
(2) a high fraction of \hbox{X-ray} detections is required to minimize 
statistical problems inherent in AGN samples with censoring, 
(3) absorbed AGNs, such as broad absorption line (BAL) quasars, 
need to be excluded when possible and controlled for when not, 
(4) radio-loud AGNs, which have an additional jet-linked \hbox{X-ray} emission 
component, need to be excluded, and 
(5) high-quality photometry/spectroscopy in the optical/UV, allowing 
mitigation of host-galaxy contamination, is required. 
We are currently completing the most detailed study to date of the
dependence of \aox\ upon redshift and luminosity, taking into account
the issues above (Strateva et~al. 2005; also see Vignali, 
Brandt, \& Schneider 2003b). We are predominantly utilizing a sample of 
156 SDSS RQQs lying serendipitously in medium-depth \rosat\ 
observations. Our partial correlation analyses 
find a highly significant (up to $10.7\sigma$) anti-correlation 
between \aox\ and 2500~\AA\ luminosity density when controlling for 
any redshift dependence (see Figure~4). In contrast, no significant 
correlation between \aox\ and redshift is found when controlling for 
the 2500~\AA\ luminosity-density dependence. Qualitatively, our results 
are consistent with much, but not all, of the earlier work on this topic. 
Quantitatively, our results are more robust and have much better control of 
statistical and systematic errors than previous studies.   

In summary, the typical RQQ (of a given luminosity) does 
not change its basic \hbox{X-ray} spectral shape or 
optical/UV-to-X-ray flux ratio over most of cosmic history. 
It appears that the small-scale \hbox{X-ray} emitting regions of RQQs are 
insensitive to the dramatic changes on larger scales that occur from 
\hbox{$z\approx 0$--6}. The data are consistent with the idea that
high-redshift RQQs are feeding and growing in basically the same
way as local ones. 


\begin{figure}[t!]
\begin{center}
\includegraphics[width=0.7\textwidth,angle=-90]{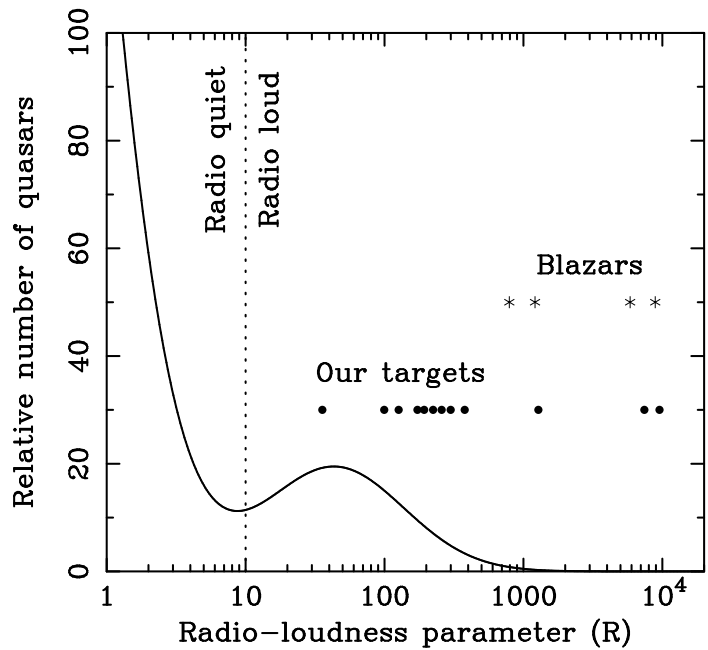}
\end{center}
\caption[]{The solid curve shows the relative number of quasars versus 
radio-loudness parameter ($R$) from Ivezi\'c et~al. (2002), and the dotted
vertical line shows the standard $R$ separation between RQQs and RLQs. 
Circles denote the $R$ values for the RLQs we have been studying, and
stars denote those for the four $z>4$ blazars with \hbox{X-ray} spectra
(the $y$-axis locations for these data points are arbitrary).}
\label{eps5}
\end{figure}


\section{High-Redshift Radio-Loud Quasars}

Until recently there were only a few $z>4$ radio-loud quasars (RLQs) 
with sensitive \hbox{X-ray} observations. Most of these were highly radio-loud 
``blazars'' with \hbox{$R\approx 1000$--10,000}; here $R$ is the standard 
radio-loudness parameter, defined as 
\hbox{$R$=$f_{\rm 5~GHz}/f_{\rm 4400~\mbox{\scriptsize\AA}}$}
(in the rest frame). While these blazars 
are not representative of the quasar population,
or even the RLQ population, as a whole (see Figure~5), they have 
been intensively targeted for \hbox{X-ray} observations due to their relatively
high \hbox{X-ray} fluxes (note the data points in Figure~1 with \hbox{0.5--2~keV}
fluxes of a \hbox{few $\times 10^{-13}$~erg~cm$^{-2}$~s$^{-1}$}). Two of the 
four $z>4$ blazars with \hbox{X-ray} spectra show evidence for substantial \hbox{X-ray} 
absorption (e.g., Worsley et~al. 2004ab), a result that is broadly 
consistent with earlier findings that the fraction of RLQs with \hbox{X-ray} 
absorption rises with redshift (e.g., Cappi et~al. 1997; Elvis et~al. 1998; 
Fiore et~al. 1998; Reeves \& Turner 2000). The observed \hbox{X-ray} absorbing gas 
is thought to be associated with the RLQs' environments, but its precise 
nature is unclear: it may be circumnuclear, located in the host galaxy, 
or entrained by the radio jets. 

We have recently performed \hbox{4--8~ks} \chandra\ snapshots to determine 
the basic \hbox{X-ray} properties of nine flat-spectrum RLQs with 
\hbox{$R\approx 30$--400} and \hbox{$z=4.0$--4.8} (Bassett et~al. 2004; 
Lopez et~al. 2005). These lie much closer than the blazars to the peak 
of the $R$ distribution for RLQs, and they are thus more representative 
of the high-redshift RLQ population (see Figure~5). We have also observed 
three recently discovered blazars with \hbox{$R\approx 1300$--9600} 
and \hbox{$z=3.5$--4.6} to enlarge the sample of these objects with \hbox{X-ray} data. 
All 12 of our targets are clearly detected, and we have performed joint 
\hbox{X-ray} spectral fitting and \aox\ analyses analogous to those described for 
RQQs in \S2. As expected from similar RLQs at low redshift, the \hbox{X-ray} 
emission from our targets usually appears to have a large contribution from 
a jet-linked spectral component, probably synchrotron self-Compton emission 
arising on sub-parsec scales. The degree of \hbox{X-ray} enhancement provided 
by this component (relative to RQQs), as well as its \hbox{X-ray} spectral shape, 
appear consistent at high and low redshift. Our analyses of \hbox{X-ray} absorption
via joint spectral fitting are ongoing, but they are challenging due to 
instrumental issues and limited photon statistics. Some \hbox{X-ray} absorption may 
be present with typical column densities of a few $\times 10^{22}$~cm$^{-2}$. 
\xmm\ spectroscopy of several of these objects is needed to establish if 
\hbox{X-ray} absorption is common among typical flat-spectrum RLQs at high redshift, 
or if it is instead confined to the minority of RLQs with $R\simgt 1000$.

Our \chandra\ observations also provide useful constraints on kpc-scale
\hbox{X-ray} jet emission from these RLQs; the on-axis angular resolution of 
\chandra\ at $z=4$ corresponds to $\approx 4$~kpc. 
One leading model for \hbox{X-ray} jet emission invokes highly relativistic 
bulk motions on kpc scales, thereby allowing electrons/positrons in the jet 
to Compton upscatter photons from the Cosmic Microwave Background (CMB) into 
the \hbox{X-ray} band (e.g., Tavecchio et~al. 2000; Celotti, Ghisellini, \& Chiaberge 2001). 
In this model, jets should remain \hbox{X-ray} bright at high redshift owing to the 
$(1+z)^4$ increase in CMB energy density which compensates for the usual
$(1+z)^{-4}$ decrease of surface brightness (Schwartz 2002; also see, 
e.g., Rees \& Setti 1968). In fact, if this model is correct, $z\simgt 4$ 
\hbox{X-ray} jets are often expected to outshine their \hbox{X-ray} cores (which dim
with luminosity distance squared). 
We have looked for such \hbox{X-ray} luminous jets in our RLQ \chandra\ observations, 
and they are not detected. Any spatially resolvable \hbox{X-ray} jets must be 
\hbox{$\approx 5$--10} times fainter than their 
\hbox{X-ray} cores.\footnote{An \hbox{X-ray} jet has been 
detected from the $z=4.30$ blazar GB~1508+5714
(Siemiginowska et~al. 2003; Yuan et~al. 2003). In this case, however, the 
jet-to-core \hbox{X-ray} flux ratio is $\approx 3$\%, substantially
less than the $\approx 100$\% discussed by Schwartz (2002).
Additionally, the putative \hbox{X-ray} jet from the $z=4.01$ blazar
GB~1713+2148 (Schwartz 2004) is not confirmed in a pointed 
\chandra\ observation (D.A. Schwartz 2004, pers. comm.).}
In some cases jets might be unresolvable 
due to an almost perfectly ``pole-on'' orientation, but such an orientation is 
unlikely for all our RLQs (\aox\ analyses also constrain this possibility). 
One plausible explanation for our nondetections is that most of the X-rays
from RLQ jets are made via a process other than CMB Compton upscattering, 
such as synchrotron radiation by multiple electron populations
(e.g., Atoyan \& Dermer 2004; Stawarz et~al. 2004). Alternatively, RLQ jets 
may often be ``frustrated'' at high redshift by their environments and thus 
only rarely achieve large angular sizes; this possibility can be tested with 
improved high-resolution radio imaging of $z>4$ RLQs. 


\begin{table}[t!]
\caption{Moderate-luminosity $z>4$ AGNs found in X-ray surveys}
\begin{center}
\renewcommand{\arraystretch}{1.4}
\setlength\tabcolsep{5pt}
\begin{tabular}{llll}
\hline\noalign{\smallskip}
AGN                         &           &  Rest-frame        &  Representative \\
name                        & Redshift  &  $\log (L_{2-10})$ &  reference      \\
\noalign{\smallskip}
\hline
\noalign{\smallskip}
CXOCY J$033716.7-050153$    &  4.61     &   44.54            &  Treister et~al. (2004)   \\ 
CLASXS J103414.33+572227    &  5.40     &   44.44            &  Steffen et~al. (2004)    \\
RX J$1052+5719$             &  4.45     &   44.72            &  Schneider et~al. (1998)  \\
CXOMP J$105655.1-034322$    &  4.05     &   44.92            &  Silverman et~al. (2005)  \\
CXOHDFN J$123647.9+620941$  &  5.19     &   44.00            &  Vignali et~al. (2002)    \\
CXOHDFN J$123719.0+621025$  &  4.14     &   43.72            &  Vignali et~al. (2002)    \\
CXOCY J$125304.0-090737$    &  4.18     &   44.39            &  Castander et~al. (2003)  \\
CXOMP J$213945.0-234655$    &  4.93     &   44.79            &  Silverman et~al. (2002)  \\
%
%
\hline
\end{tabular}
\end{center}
The third column above is the rest-frame \hbox{2--10~keV} luminosity
(in erg~s$^{-1}$), computed using a power-law photon index of $\Gamma=2$. 
We have only included AGNs in this table with $\log (L_{2-10})<45$. A few 
higher luminosity AGNs have also been found in \hbox{X-ray} surveys, such as 
RX~J$1028.6-0844$ (Zickgraf et~al. 1997) and 
RX~J$1759.4+6638$ (Henry et~al. 1994). 
\label{table1}
\end{table}


\section{X-ray Survey Constraints on Moderate-Luminosity
Active Galactic Nuclei at High Redshift}

The analyses above engender confidence that \hbox{X-ray} selection should remain 
effective at finding AGNs at the highest redshifts, and the deepest current
surveys with \chandra\ and \xmm\ (e.g., Brandt \& Hasinger 2005) have 
sufficient sensitivity to detect $z>4$ AGNs that are $\approx 10$--30 times 
less luminous than the quasars found in wide-field optical surveys
(see Figure~1). For example, AGNs similar to 
Seyfert galaxies in the local universe, with 
\hbox{X-ray} luminosities of $\simgt 5\times 10^{43}$~erg~s$^{-1}$, should be 
detectable to $z\approx 10$ in the 2~Ms \chandra\ Deep Field-North (CDF-N; 
Alexander et~al. 2003) observation. Such moderate-luminosity AGNs are much 
more numerous and thus more representative of the AGN population than the 
luminous quasars from wide-field optical surveys. Furthermore, \hbox{X-ray} 
surveys suffer from progressively less absorption bias as higher redshifts are 
surveyed (penetrating $\approx 2$--40~keV rest-frame \hbox{X-rays} are 
sampled at $z>4$), whereas optical surveys sample rest-frame ultraviolet 
light which can be absorbed by dust. 

Moderate-luminosity \hbox{X-ray} detected AGNs at \hbox{$z=4$--6.5} are expected 
to have $I$-band magnitudes of \hbox{$\approx 23$--27}; optical spectroscopy 
of these objects is thus challenging. Nevertheless, significant constraints on 
sky density have been set via intensive follow-up studies with large optical
telescopes and color selection (e.g., Alexander et~al. 2001; Barger et~al. 2003; 
Cristiani et~al. 2004; Koekemoer et~al. 2004; Wang et~al. 2004b). The sky density 
of $z>4$ AGNs is \hbox{$\approx 30$--150~deg$^{-2}$} at a \hbox{0.5--2~keV} flux 
limit of $\approx 10^{-16}$~erg~cm$^{-2}$~s$^{-1}$; for comparison, the 
sky density of $z=4$--5.4 SDSS quasars is $\approx 0.12$~deg$^{-2}$ at an 
$i$-magnitude limit of $\approx 20.2$ (e.g., Schneider et~al. 2003).
While the constraints on the faint end of the AGN \hbox{X-ray} luminosity 
function at \hbox{$z>4$} have large statistical errors, the data are already
sufficient to argue convincingly that the AGN contribution to reionization 
at $z\approx 6$ is small. 


\begin{figure}[t!]
\begin{center}
\includegraphics[width=1.0\textwidth,angle=0]{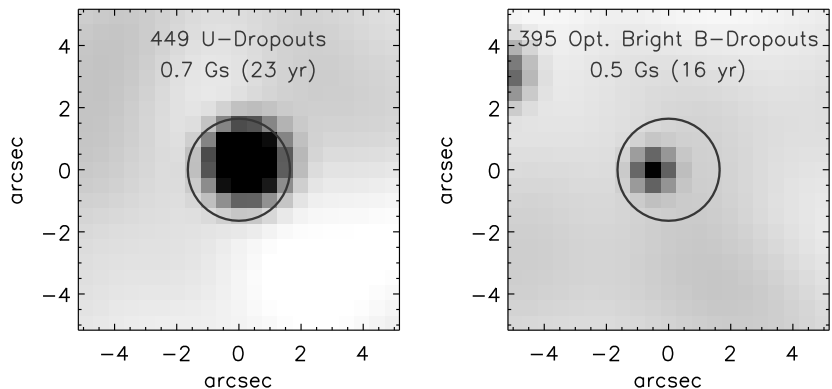}
\end{center}
\caption[]{Stacked images of Lyman break galaxies from the Great Observatories 
Origins Deep Survey (GOODS) in the 0.5--2~keV band. The left panel shows the 
stacking results for 449 $U$-dropouts with a typical redshift of $z=3.0$
($7.1\sigma$ detection), and the right panel shows those for 395 
optically bright $B$-dropouts with a typical redshift of $z=3.8$
($3.2\sigma$ detection). The effective exposure times for the left
and right panels are 23 and 16~yr, respectively. Adapted from 
Lehmer et~al. (2005).}
\label{eps6}
\end{figure}


Table~1 lists the moderate-luminosity \hbox{$z>4$} AGNs found in \hbox{X-ray}
surveys. The number of objects is small but is increasing fairly 
rapidly, as can be seen from the publication dates of the associated
papers. With appropriate effort it should be possible to generate
$\approx 50$ moderate-luminosity \hbox{$z>4$} AGNs over the next 
$\approx 5$ years, thereby defining in respectable detail the faint 
end of the luminosity function. Current \hbox{X-ray} spectral and \aox\ 
analyses, albeit limited, suggest that the moderate-luminosity 
\hbox{$z>4$} AGNs found in \hbox{X-ray} surveys have similar basic 
emission properties to comparably luminous objects at low
redshift (e.g., Vignali et~al. 2002). 

High-redshift AGN populations with even lower luminosities can be constrained 
with \hbox{X-ray} source-stacking analyses. These search for an average 
signal from a set of high-redshift sources whose individual members lie below 
the single-source \hbox{X-ray} detection limit. The most sensitive \hbox{X-ray} 
source-stacking analyses at $z\approx 4$ or higher have employed samples of 
$\approx 250$--1700 Lyman break galaxies (e.g., Lehmer et~al. 2005; 
$B$, $V$, and $i$ dropouts) and $\approx 100$ Ly$\alpha$ emitters 
(e.g., Wang et~al. 2004a; at $z\approx 4.5$). Average \hbox{X-ray} detections have 
presently been obtained up to $z\approx 4$ (see Figure~6), while physically 
interesting upper limits are obtained at higher redshifts. The data are 
plausibly consistent with any \hbox{X-ray} emission from these objects 
arising from stellar processes (e.g., \hbox{X-ray} binaries and supernova
remnants); \hbox{X-ray} emission from numerous, low-luminosity 
AGNs is not required. These average constraints at luminosities below those 
that can be probed by single-source analyses further limit the contribution 
that AGNs could have made to reionization at $z\approx 6$. A complementary 
average constraint, derived by considering the unresolved component of the 
\hbox{0.5--2~keV} background, provides additional evidence that AGNs 
and lower mass black holes did not dominate reionization 
(Dijkstra, Haiman, \& Loeb 2004). 


\section{Some Future Prospects}

The work described above can be extended in several ways with 
\chandra\ and \xmm\ observations: 

\begin{enumerate}

\item
The number of \hbox{X-ray} detections at $z>5$ is still relatively small
at present, and this can be increased substantially with appropriate
\chandra\ snapshots of the highest redshift quasars. 

\item
Only the most basic \hbox{X-ray} properties of typical RLQs at $z>4$ are now known. 
\xmm\ spectroscopy should allow their \hbox{X-ray} continuum and absorption 
properties to be determined much better, and deeper \chandra\ imaging of some
objects is needed to search for \hbox{X-ray} jets more sensitively (in conjunction
with improved radio imaging). Steep-spectrum RLQs at $z>4$ also need to be
studied in the \hbox{X-ray} band. 

\item
Additional \hbox{X-ray} investigations of minority AGN populations at $z>4$, 
such as weak emission-line quasars and BAL quasars, can provide insight into 
their nature. 

\item
X-ray studies of $z>4$ AGNs selected at infrared, submillimeter, and millimeter 
wavelengths (or with remarkable properties at these wavelengths) can 
establish the relative importance of AGN vs. stellar processes in these
objects and allow assessment of extinction biases. 

\item
The number of known moderate-luminosity AGNs at $z>4$ should increase
significantly over the next few years, as follow-up studies of the many
ongoing \hbox{X-ray} surveys progress. Uniform searches for 
such AGNs are particularly important for setting reliable demographic 
constraints. 

\item
Deeper \hbox{X-ray} surveys than those to date (e.g., \hbox{5--10~Ms} \chandra\ 
observations) would allow improved searches for highly obscured AGNs at 
$z>4$. 

\item
X-ray stacking analyses can be improved as high-redshift source samples 
are enlarged and refined. 

\end{enumerate}

\noindent
In the more distant future, \conx\ and \xeus\ should allow \hbox{X-ray} 
spectroscopy down to \hbox{0.5--2~keV} flux levels of 
\hbox{$\approx 5\times 10^{-16}$} and 
\hbox{$\approx 5\times 10^{-17}$~erg~cm$^{-2}$~s$^{-1}$}, 
respectively (see Figure~1). Precise \hbox{X-ray} measurements of continuum shape, 
absorption, iron~K line emission, and Compton reflection should be possible
for many $z>4$ AGNs. Ultimately, a mission such as \genx\ should be able to 
survey the \hbox{$z\approx 10$--15} \hbox{X-ray} universe, searching for X-rays
from proto-quasars and the $\approx 100$--300~M$_\odot$ black
holes made by the deaths of the first stars. 


\section*{Acknowledgments}

We thank all of our collaborators on this work. 
We thank D.M. Alexander, L.C. Bassett, F.E. Bauer, 
C.C. Cheung, S. Heinz, D.A. Schwartz, and 
A.T. Steffen for helpful discussions. 
We acknowledge funding from 
NASA LTSA grant NAG5-13035 (WNB, DPS, IVS), 
NSF CAREER award AST-9983783 (WNB, BDL), 
\chandra\ and \xmm\ grants (WNB, DPS), and
MIUR COFIN grant 03-02-23 (CV). 



\end{document}